\documentclass[conference]{IEEEtran}
\IEEEoverridecommandlockouts
\usepackage{hyperref}
\usepackage{cite}
\usepackage[ruled,vlined]{algorithm2e}
\usepackage{booktabs}
\usepackage{amsmath,amssymb,amsfonts}
\usepackage{enumitem} 
\usepackage{algorithmic}
\usepackage{graphicx}
\usepackage{textcomp}
\usepackage{xcolor}
\def\BibTeX{{\rm B\kern-.05em{\sc i\kern-.025em b}\kern-.08em
    T\kern-.1667em\lower.7ex\hbox{E}\kern-.125emX}}
\begin{document}

\title{EdgeMLBalancer: A Self-Adaptive Approach for Dynamic Model Switching on Resource-Constrained Edge Devices\\}

\author{
    \IEEEauthorblockN{
        Akhila Matathammal\IEEEauthorrefmark{1}, 
        Kriti Gupta\IEEEauthorrefmark{2}, 
        Larissa Lavanya\IEEEauthorrefmark{2},\\
        Ananya Vishal Halgatti\IEEEauthorrefmark{2},         
        Priyanshi Gupta\IEEEauthorrefmark{2}, 
        Karthik Vaidhyanathan\IEEEauthorrefmark{1}
    }
    \IEEEauthorblockA{\IEEEauthorrefmark{1}Software Engineering Research Center, IIIT Hyderabad, India \\
    \{akhila.matathammal@research.iiit.ac.in, karthik.vaidhyanathan@iiit.ac.in\}}
    \IEEEauthorblockA{\IEEEauthorrefmark{2}IIIT Hyderabad, India \\
    \{kriti.g, larissa.lavanya, ananya.halgatti, priyanshi.gupta\}@students.iiit.ac.in}
}

\maketitle

\begin{abstract}
\noindent
The widespread adoption of machine learning (ML) on edge devices, such as mobile phones, laptops, IoT devices, etc., has enabled real-time AI applications in resource-constrained environments. Existing solutions for managing computational resources often focus narrowly on accuracy or energy efficiency, failing to adapt dynamically to varying workloads. Furthermore, the existing system lack robust mechanisms to adaptively balance CPU utilization, leading to inefficiencies in resource-constrained scenarios such as real-time traffic monitoring. To address these limitations, we propose a Self-Adaptive approach that optimizes CPU utilization and resource management on edge devices.  

Our approach  EdgeMLBalancer balances between models through dynamic switching, guided by real-time CPU usage monitoring across processor cores. Tested on real-time traffic data, the approach adapts object detection models based on CPU usage, ensuring efficient resource utilization. The approach leverages epsilon-greedy strategy which promotes fairness and prevents resource starvation, maintaining system robustness. The results of our evaluation demonstrate significant improvements by balancing computational efficiency and accuracy, highlighting the approach’s ability to adapt seamlessly to varying workloads. This work lays the groundwork for further advancements in self-adaptation for resource-constrained environments.  
\end{abstract}


\section{Introduction}

Artificial Intelligence (AI) is increasingly integrated into everyday technologies, with smartphones serving as prominent edge devices where many machine learning (ML) functionalities are embedded \cite{b1}.  These devices now support real-time ML applications, leveraging proximity to data sources for improved privacy and reduced latency. However, deploying the computationally intensive ML models on resource-constrained devices introduces critical challenges, including limited memory, energy storage, and processing power \cite{b2}.  

While lightweight and quantized models offer solutions to reduce resource demands, they often compromise on accuracy \cite{b14}\cite{b15}, which is critical for applications such as autonomous vehicles and augmented reality, these optimized models cannot fully adapt to varying workloads or operational contexts, where real-time responsiveness and precision are simultaneously required \cite{b16}\cite{b17}\cite{b18}. The challenge lies not just in model selection but in dynamically balancing trade-offs between computational efficiency and accuracy based on situational demands \cite{b3}. Unlike cloud systems, where scalability and abundant resources mitigate such constraints, edge devices must operate within finite computational capacities. This limitation underscores the critical need for efficient and adaptive resource management techniques that balance precision and performance, particularly for real-time applications \cite{b4}. Significant advances have been made to address these issues. Techniques such as dynamic model switching, TinyML, and neural network compression optimize resource utilization while maintaining system precision \cite{b5}\cite{b6}. Frameworks such as EcoMLS, Ada-HAR have demonstrated energy-efficient solutions for managing workloads~\cite{b7}\cite{b8}.

However, the majority of existing approaches focus on cloud-based or Mobile Edge Computing (MEC) systems \cite{b19}\cite{b20}\cite{b21}, where resource constraints are mitigated through offloading computations to cloud servers or nearby edge infrastructures. These methods excel in leveraging shared resources but are not directly applicable to scenarios involving standalone edge devices, such as smartphones, which operate independently without external computational support. Additionally, many approaches emphasize static optimization strategies or cater to specific applications, leaving room for dynamic, adaptable frameworks that can address the diverse and evolving requirements of real-world edge applications. 

Self-adaptive systems have emerged as a promising solution to handle run-time uncertainities, making them particularly valuable in resource-constrained settings where resources are limited and conditions are uncertain. These systems leverage feedback loops to continuously monitor, analyze, and adjust their behavior to meet predefined goals \cite{b35}. For instance, probabilistic modeling frameworks evaluated the cost-benefit trade-offs of adaptation actions, ensuring that system utility is optimized without compromising performance \cite{b23}. Although adaptive techniques such as QoS-aware model switching and modular software designs \cite{b3}\cite{b15} have shown promise, they are often tailored to specific applications and primarily focus on cloud-based systems. These approaches address trade-offs between computational efficiency and accuracy but are not directly applicable to scenarios involving real-time constraints on resource-limited devices such as smartphones. 

To address these challenges, this study introduces a dynamic model-switching approach, EdgeMLBalancer for object detection on edge devices. Building on the principles of self-adaptation, the concept of ML Balancer is introduced in this study. The ML Balancer is designed to dynamically evaluate and select between ML models based on accuracy and resource demands. The approach is based on an epsilon-greedy strategy to promote fairness and to prevent model starvation. Given a real-time traffic monitoring scenario, our approach (i) continuously monitors CPU usage and accuracy, (ii) selects the most suitable model adaptively, balancing computational efficiency and accuracy, (iii) dynamically switches between models based on epsilon-greedy decision-making strategy to facilitate fairness and responsiveness, and (iv) prevents overutilization of specific models by distributing workloads across available options. The EdgeMLBalancer approach is prototyped on \textit{Qualcomm QIDK platform}\footnote{\url{https://www.qualcomm.com/products/mobile/snapdragon/smartphones/snapdragon-8-series-mobile-platforms/snapdragon-8-gen-2-mobile-platform}}, to simulate different scenarios, and is further evaluated on real-time traffic data using the edge-device (smartphone). The evaluation demonstrates significant improvements in balancing computational efficiency and accuracy, facilitating fairness of model usage compared to other approaches, validating effectiveness of the proposed EdgeMLBalancer approach in achieving optimzal performance under varying runtime conditions.

The remainder of the paper is structured as follows: Section II provides a running example. Section III introduces the our approach. Experimentation and results from its application are in Section IV. Threats to validity and Related work are discussed in Section V and VI respectively. Section VII concludes and discusses future work.

\section{Running Example}

\begin{figure}
    \centering
    \includegraphics[width=1\linewidth]{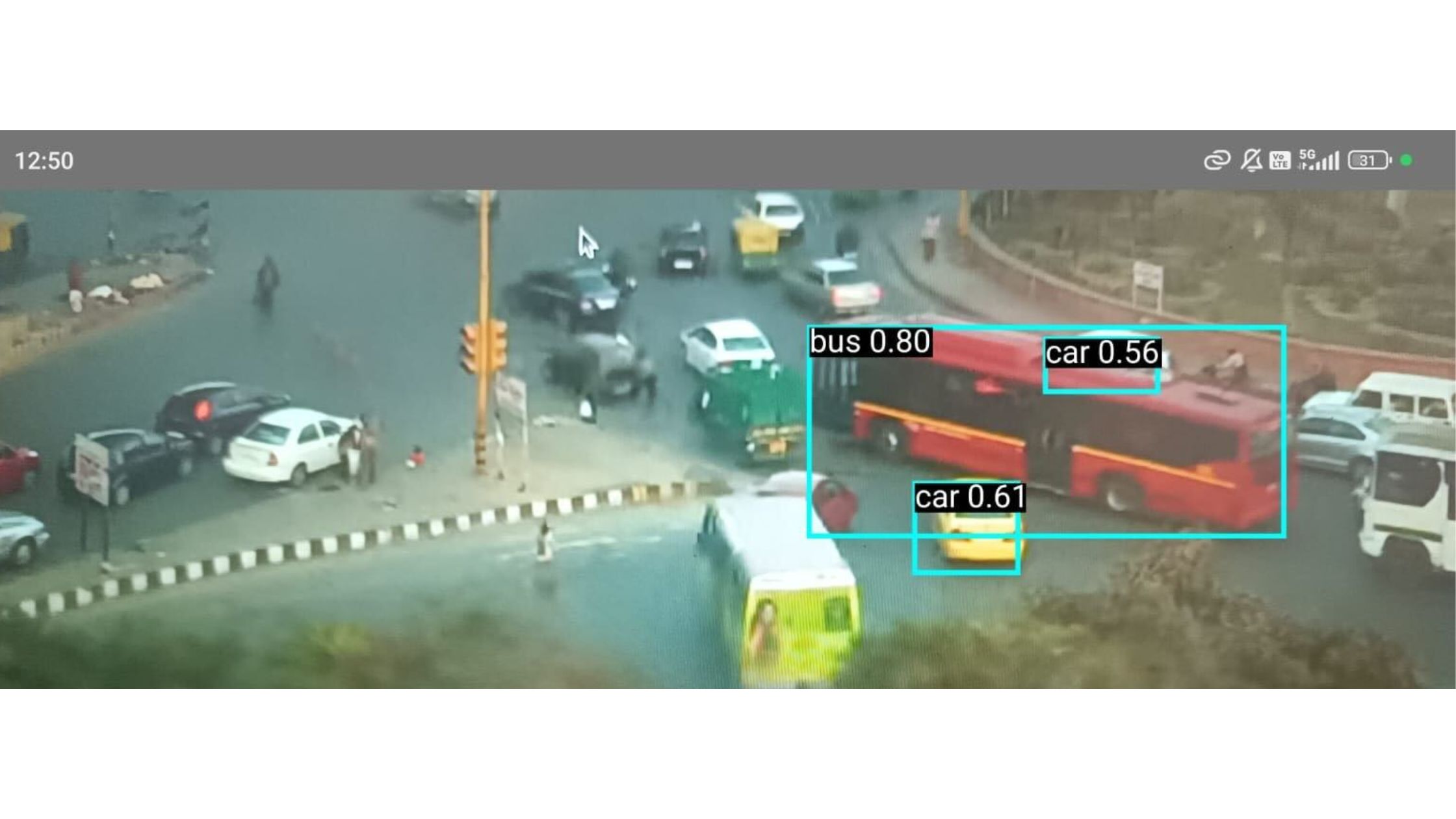}
    \caption{Real-Time Object Detection Example Frame}
    \label{fig:running-example}
\end{figure}

We illustrate our EdgeMLBalancer approach on a scenario that concerns real-time traffic data analysis for object detection on Indian roads, chosen for their diverse and complex traffic conditions. 
Rather than relying on a single model, which may perform well under specific conditions but fails to adapt to dynamic changes in traffic density or vehicle behavior, self-adaptive systems dynamically adjust model selection based on runtime conditions. This adaptability is particularly critical in environments such as Indian roads, where traffic patterns are highly variable due to factors such as time of day, location, and road conditions. For instance, traffic density can peak during rush hours and significantly drop during off-peak times. A fixed-model approach may either overutilize system resources during low-traffic periods or fail to maintain the required accuracy during high-density periods, leaving to inefficiencies in both performance and resource usage.

In such scenarios, the choice of edge devices over centralized cloud systems becomes essential. Edge devices enable real-time inference directly on the device, local processing of data, minimizing latency and reducing reliance on stable internet connectivity-factors that are particularly important in regions with inconsistent network infrastructure, unlike cloud systems that add transmission delays and high energy and operational costs. 

The EdgeMLBalancer system uses the \textit{Image Capture Module} as shown in \textit{Figure}\ref{fig:architecture}, where a smartphone camera streams real-time traffic frames such as in \textit{Figure}\ref{fig:running-example}, emulating real-world scenarios. These frames are then passed to the \textit{Pre-processing Module}, which resizes and normalizes the data to meet the requirements of collection of models deployed on edge where each model \(m_i\) in \textit{M} represents different configurations, including EfficientDet Lite0, Lite1, Lite2, and SSD MobileNet V1 \cite{b36}\cite{b37}. Some of the models offer high accuracy demanding more computational power while other demanding less with moderate accuracy. The preprocessed frames are processed in the \textit{Object Detection} component, where the system processed each frame by selecting a model based on the trade-off between real-time CPU usage, and inference accuracy. Finally, the \textit{Post-Processing Module} refines detection results by filtering low-confidence predictions and overlays bounding boxes and confidence scores onto the original frames for visualization, as shown in \textit{Figure}\ref{fig:running-example}. The entire system effectively demonstrates the feasibility of self-adaptive systems in real-time applications, to balance trade-off between efficient resource utilization and accuracy, making it a sustainable and robust solution for resource-constrained edge devices. 

\section{EdgeMLBalancer Approach}

\begin{figure*}[ht]
    \centering
    \includegraphics[width=1\linewidth]{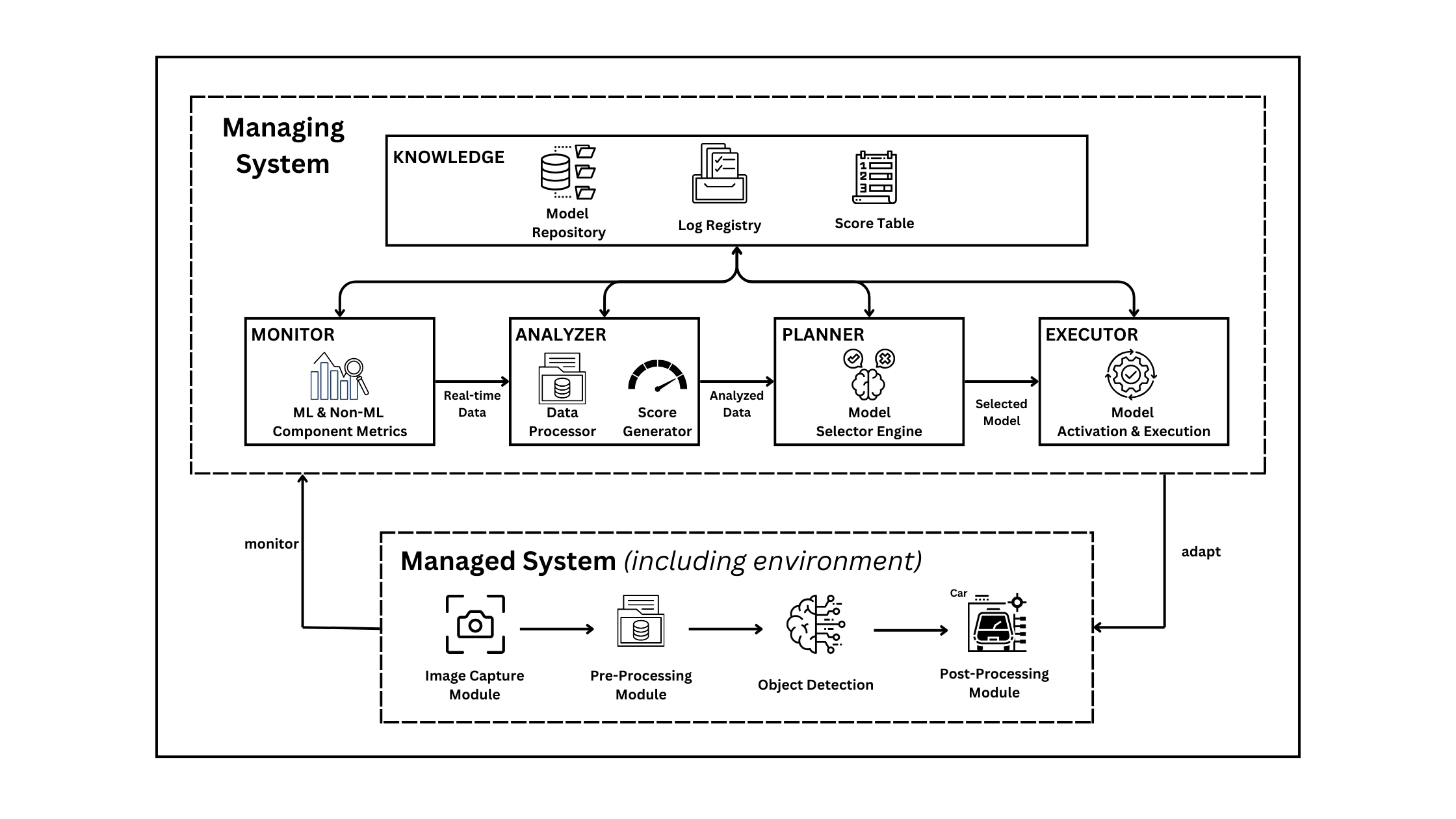}
    \caption{Architecture of EdgeMLBalancer}
    \label{fig:architecture}
\end{figure*}

We use the running example presented in the Section II to explain the approach. The goal of our EdgeMLBalancer approach is to enable efficient and adaptive real-time object detection on resource-constrained edge devices. By leveraging self-adaptation, our approach dynamically adjusts system behavior, such as model selection based on real-time CPU usage, to optimize resource utilization while maintaining high inference accuracy, as shown in \textit{Figure} \ref{fig:architecture}.

To achieve this, our system leverages a modular architecture. In the rest of this section, we will explain our approach using the running example presented in Section II. The \textit{Managing System} is at the core of this architecture, relying on the MAPE-K loop \cite{b13} to monitor runtime metrics, analyze performance, and select models based on CPU usage and accuracy trade-offs using an epsilon-greedy strategy. This design allows for a dynamic adaptation to changing environmental conditions, aiming to achieve a trade-off to balance performance, energy (through CPU), and inference accuracy.

\subsection{MAPE-K Feedback Loop}

The approach makes use of the Monitor-Analyze-Plan-Execute-Knowledge (MAPE-K) loop, enabling continuous monitoring, decision-making, and adaptation based on real-time data. Below, each component of the MAPE-K loop is described in detail.

\subsubsection{\textbf{Monitor}}

The \textit{monitoring component} operates as a data aggregation and reporting layer, ensuring the \textit{Managing System} has access to the latest operational information about the system and its environment. The metrics collected by the \textit{Monitor} are categorized into two groups as shown in \textit{Figure} \ref{fig:architecture}: \textit{ML Component Metrics} and \textit{Non-ML Component Metrics}. 

\noindent
\textbf{ML Component Metrics:}

It focuses on metrics directly related to the performance of the object detection models (can be extended to other classes of models as well). The metrics collected are:

\begin{itemize}
    \item Confidence Score (\textit{C}): Measures the reliability of the model's predictions for each object of the given frame. For a frame \textit{i} with \(d\) detected objects, the confidence score \(C_i\) is computed as: 
    \[C_i = \frac{\sum_{j=1}^{d} c_j}{d}
    \] 
    For instance, if a frame contains three detections with confidence scores of 0.85, 0.75, and 0.9, then the overall confidence score for that frame is computed using the above formula as 0.833. This score helps evaluate the model's accuracy and stability, where stability refers to the model's ability  consistently to produce reliable predictions across varying conditions and workloads. 

\end{itemize}

Apart from the Confidence score (which is the main metric used in the context of this study), other metrics that can be monitored include the number of detections, model size, throughput (in terms of detections per unit of time), etc.

\noindent
\textbf{Non-ML Component Metrics:}

It captures metrics related to the system resource usage, and data from the managed system and its environment, emphasizing the computational constraints of edge devices. The metrics collected are:

\begin{itemize}
    \item CPU Usage (\textit{U}): Monitors percentage of processor utilization for model inference performed on each frame, providing real-time insights into the computational load imposed by the active model. High CPU usage can indicate bottlenecks, necessitating a switch to a less resource-intensive model. 

\end{itemize}

The metrics measured are focused on our study and this can be extended to other metrics like measuring memory usage, battery usage, network utilization, etc.

The set of monitored metrics Metrics (\(M_\text{data}\)) capturing system resource usage \(U_i\), model performance \(C_i\), and the current model in use \(m_i\). For each frame \textit{i}, these metrics are collected continuously, allowing the system to maintain real-time awareness of both model performance and resource utilization. 
\begin{align*}
M_{\text{data}} = \{(m_i, C_i, U_i) | i = 1,2,...,n\}
\end{align*}
\textit{Example} In the running example section of traffic detection, \(M_\text{data}\) for frame \(i\) might include a CPU usage of \(13\%\), an inference accuracy of \(54.42\%\) for \(3\) detected objects, and the model name "EfficientDet Lite2" (refer Section II).

This granular real-time monitoring provides the basis for the \textit{Analyzer} to evaluate trends and adapt the system dynamically. This component not only monitors but also logs the metrics data into the \textit{Log Repository} in \textit{Knowledge} (discussed in later part of this section), facilitating a thorough performance review. 

\subsubsection{\textbf{Analyzer}}

It plays a critical role in assessing the performance of the current model and computing scores that guide model selection in the \textit{Planner}. The \textit{Analyze} component consists of two submodules as shown in \textit{Figure} \ref{fig:architecture}: the \textit{Data Preprocessor} and the \textit{Score Generator}.

The \textit{Data Preprocessor} begins by cleaning and organizing the raw data gathered from \textit{Monitor}. For each frame \textit{i}, metrics such as confidence score (\(C_i\)), and CPU usage (\(U_i\)) are collected and then the aggregation is performed over the window of last \textit{n} frames, to capture overall performance trends, with the average values \(C_\text{avg}, U_\text{avg}\) calculated as:
\[C_\text{avg} = \frac{1}{n} \sum_{i=1}^{n} C_i \quad 
U_\text{avg} = \frac{1}{n} \sum_{i=1}^{n} U_i\]

The aggregated metrics for each model, represented as (\(C_\text{avg}, U_\text{avg}\)), are then structured into a preprocessed dataset, which forms as input for the \textit{Score Generator}.

\textit{Example} Considering the data from the example mentioned in the \textit{Monitor component} and with historical data, the aggregated data for the frame is calculated to be \(U_\text{avg} = 18\%\), and \(C_\text{avg} = 55.94\%\). Now this aggregated data forms as an input to the score generator.

The \textit{Score Generator} evaluates the preprocessed data to compute a performance score (\(S\)) \cite{b5} for each model. This score combines real-time (\(C_i, U_i\)) and historical metrics (\(U_\text{avg}, C_\text{avg}\)) to evaluate the trade-offs between computational efficiency and detection accuracy. The performance score \(S_{m_i}\) for a model \(m_i\) is calculated as:
\[
S_{m_i} = \arg\min(U_i, U_\text{avg}) \times (1 - \frac{C_\text{avg}}{C_i})
\]
The score calculation serves two purposes. One, \textit{efficiency assessment}, minimizing the CPU usage term (\(\min(U_i, U_\text{avg})\)), the system favors models with lower computational demands. Other, \textit{accuracy assessment}, where the confidence ratio (\((1 - \frac{C_\text{avg}}{C_i})\)) highlights models that are maintaining or improving their accuracy relative to historical performance. 

\textit{Example} Assume \( U_\text{avg}, C_\text{avg}\) values from the \textit{data preprocessor example} are the inputs for historical data, and  \(U_i, C_i\) are the inputs for current values. Now using the \(S_{m_i}\) formula, the computed score for the "EfficientDet Lite2" model is \(S_{m_i} = -0.3627\). The negative score reflects slight decline in the confidence, but the low CPU usage reduces the penalty, making it a better score. 

This \textit{Score} is then given as an input to the \textit{Planner} component and also logs this data into the \textit{Score Table} in \textit{Knowledge} component (discussed later in this section). The \textit{Analyze} component transforms raw runtime data into actionable performance score, bridging the gap between monitoring and planning.

\subsubsection{\textbf{Planner}}

This component in our approach serves as the decision-making hub responsible for selecting the optimal model for inference. It uses the performance scores generated in the \textit{Analyze} component and determines which model will be deployed next. By leveraging the \textit{Epsilon-Greedy Strategy}, the \textit{Plan} component balances the need for exploration (testing alternative models) and exploitation (using the best-performing model) to trade-off between adaptability, efficiency, and fairness. 

\noindent
\textbf{Model Selector Engine:}

At the core of the \textit{Plan} component is the \textit{Model Selector Engine} as shown in \textit{Figure} \ref{fig:architecture}, which evaluates all available models stored in the \textit{Score Table}. The engine considers the performance scores for each model, reflecting their performance under the current operational conditions. The decision-making process is guided by the \textit{Epsilon-Greedy Strategy}, ensuring that no model is starved and that system performance is continually optimized. The \textit{Model Selector Engine} operates as follows:

\begin{enumerate}
    \item Retrieves performance scores \(S\) for all models from the Score Table.
    \item Applies Epsilon-Greedy Strategy (discussed later in this component), to select the next model for inference.
    \item Updates the \textit{Executor} component with the selected model and its operational context. 
\end{enumerate}

\noindent
\textbf{Epsilon-Greedy Strategy:}

This is a well-known probabilistic decision-making mechanism, primarily used in adaptive systems, to balance \textit{exploration \& exploitation} effectively \cite{b38}\cite{b39}. Its core idea is to ensure the system does not rely exclusively on the currently best-performing model (exploitation) but also occasionally tests alternative models (exploration) to discover potentially better options as system conditions evolve. In the running example section of traffic detection on edge devices, where workloads, CPU availability, and environmental factors fluctuate, epsilon-greedy enables adaptability by periodically switching models that may perform better under new conditions. This strategy presented in the \textit{Algorithm}\ref{alg:selection}, ensures fairness and prevents overutilization of any single model, balances performance with long-term adaptability, and minimizes overhead, making it ideal for resource-constrained environments. It operates as follows: 

\begin{enumerate}
    \item For each inference cycle, a random number is generated as shown in \textit{Algorithm} \ref{alg:selection} (line 4).
    \item Exploration: With a probability \(\epsilon\), the system selects a model randomly from the available repository, as in \textit{Algorithm} \ref{alg:selection} [lines 5-6]. This exploration step evaluates underutilized models and updates their performance metrics. 
    \item Exploitation: With a probability \(1-\epsilon\), the system selects the model with the lowest performance score from the \textit{Score Table}, as in \textit{Algorithm} \ref{alg:selection} [lines 7-11].
\end{enumerate}

\noindent
\textit{Example} In the running example section of traffic detection, given the performance scores for the available 4 models as \(S_\text{efficientdet-lite0} = -0.25, S_\text{efficientdet-lite1} = -0.30, S_\text{efficientdet-lite2} = -0.36, S_\text{ssd-mobilenet-v1} = -0.28\), the random value \(p\) determines the model selection in \textit{Algorithm} \ref{alg:selection}[line 4]. If \(p = 0.3\), then the algorithm executes exploitation [lines 7-12], and selecting \(m_\text{selected} = \text{"EfficientDet Lite2"}\) in [line 14], as it has the lowest score. Conversely, if \(p = 0.08\), then algorithm executes exploration [lines 5-6], randomly selecting \(m_\text{selected} = \text{"EfficientDet Lite0"}\) in [line 14].

The \textit{Plan} component concludes by forwarding the selected model to the \textit{Executor} component for inference. The epsilon-greedy being a light-weight algorithm, allows you to easily select and dynamically adapt to evolving workloads and operational conditions, optimizing performance while maintaining resource efficiency. 

\subsubsection{\textbf{Executor}}

The \textit{Execute} component of the MAPE-K loop is responsible for enacting the decisions made in the \textit{Plan} component by deploying the selected model for inference. This component ensures seamless integration between model selection and real-time object detection, maintaining system responsiveness and adaptability. The \textit{Execute} component performs two key functions: \textit{Model Activation and Execution} and \textit{Feedback Logging}.

Upon receiving the selected model (\(m_\text{selected}\)) from the \textit{Plan} component as in \textit{Algorithm} \ref{alg:selection} [line 14], the \textit{Execute} component activates and loads the corresponding TensorFlow Lite model from the \textit{Model Repository} in \textit{Knowledge}. The selected model processes each incoming frame by performing \textit{Object Detection} and \textit{Post-Processing} as explained in the running example section. 

If the selected model differs from the previously active model (\(m_\text{active} \neq m_\text{selected}\)), the currently active model is deactivated to free resources, and the new model is initialized. This ensures smooth transitions during model switches maintaining system responsivesness and reliability. 

Following each inference, the \textit{Execute} component logs metrics (\(M_\text{data}\)) into the \textit{Knowledge Log Repository}. These logs serve as updated inputs for the Monitor, enabling the system to continuously adapt to changing operational conditions while refining model selection in future cycles by feedback logging. 

\subsubsection{\textbf{Knowledge}}

The \textit{Knowledge} component serves as the central repository, storing critical data required for adaptive decision-making. It supports all the components by maintaining up-to-date records of models, performance scores, operational logs, ensuring informed and efficient system adaptation. It consists of three submodules as shown in \textit{Figure} \ref{fig:architecture}: the \textit{Model Repository}, the \textit{Score Table}, and \textit{Log Registry}. 

The \textit{Model Repository}  houses a variety of preloaded object detection models \textit{M}, where each model \(m_i\)  in \textit{M} represents different configurations, including EfficientDet Lite0, Lite1, Lite2, and SSD MobileNet V1 \cite{b36}\cite{b37}. Each model represents a unique trade-off between computational efficiency and detection accuracy. For instance, SSD MobileNetV1 is generally more suitable for scenarios requiring low latency, and limited computational resources, offering faster inference times with moderate accuracy [9][10]. In contrast, EfficientDet Lite0 provides higher accuracy but demands more computational power, leading to increased CPU usage and energy consumption [9][10].  This enables quick model initialization during execution and supports model comparison during analysis. 

The \textit{Score Table} maintains the performance scores for each model, dynamically updated after every analysis cycle. These scores guide the \textit{Planner} in selecting the optimal model. Additionally, the \textit{Log Registry} records runtime metrics such as CPU usage, battery usage, confidence scores, and number of detections, offering historical data for performance trend analysis. It plays a vital role in enabling efficient model selection, trend analysis, and runtime  updates, making it cornerstone of the self-adaptive framework. 

\textit{Example} In the running example section of traffic detection, all the 4 available models and its configurations are stored in \textit{Model Repository}, enabling the system to quickly load and evaluate models when required. The given performance scores computed in \textit{score generator example} and forwarded to \textit{planner example} are stored and dynamically updated in the \textit{Score Table}, based on which we select the model if it is exploitation in \textit{Planner} component. \textit{Log Registry} records the runtime metrics for historical analysis.

\begin{algorithm}
\begin{small}
\caption{Planner: Algorithm for Model Selection with Epsilon-Greedy}\label{alg:selection}
\begin{algorithmic}[1]
\STATE \textbf{procedure} FORMULATOR(\(m_\text{active}\), \(U_\text{active}\), \(C_\text{active}\)) \(\triangleright\) Input: Current active model (\(m_\text{active}\)), its CPU usage (\(U_\text{active}\)), and its inference confidence (\(C_\text{active}\))
\STATE \textbf{Initialize:}
\STATE \hspace{1em} $M \gets \{ m_1, m_2, \dots, m_n \}$
\STATE \hspace{1em} $p \sim \textit{random}(0, 1)$

\IF{ \(p \leq \epsilon \textbf{ then}) \)\(\triangleright\) Exploration: Randomly select model excluding the best model }
    \STATE \(M_{\text{next}} \gets random(M)\)
\ELSE 
    \STATE \(\triangleright\) Exploitation: Select the best model based on Score
    \FOR{each \(m_{\text{i}} \in M\)}
        \STATE Get Calculated Score from Analyzer component
        \STATE \(M_{\text{next}} \gets \arg\min(S)\)
    \ENDFOR
\ENDIF

\STATE Return selected model: \(m_\text{selected}\)
\end{algorithmic}
\end{small}
\end{algorithm}

\section{Experimentation and Results}

\begin{table*}[ht]
    \centering
    \resizebox{\textwidth}{!}{%
    \begin{tabular}{@{}|l|c|c|c|c|c|c|@{}}
        \toprule
        \textbf{Approach} & \textbf{Frames Processed} & \textbf{Average CPU Usage (\%)} & \textbf{Average Accuracy (\%)} & \textbf{Average Switching Time (s)} & \textbf{Battery Consumption (mAh)} \\ 
        \midrule
        Epsilon-Greedy & 1952 & 19.90 & 17.36 & 0.85 & 2.10 \\ 
        Naive & 2482 & 20.63 & 2.94 & 0.50 & 5.25 \\ 
        Round Robin With Boosting & 2458 & 19.08 & 10.85 & 1.40 & 3.10 \\ 
        \bottomrule
    \end{tabular}%
    }
    \caption{Performance metrics comparison of different approaches.}
    \label{tab:table}
\end{table*}

\begin{figure}
    \centering
\includegraphics[width=\linewidth]{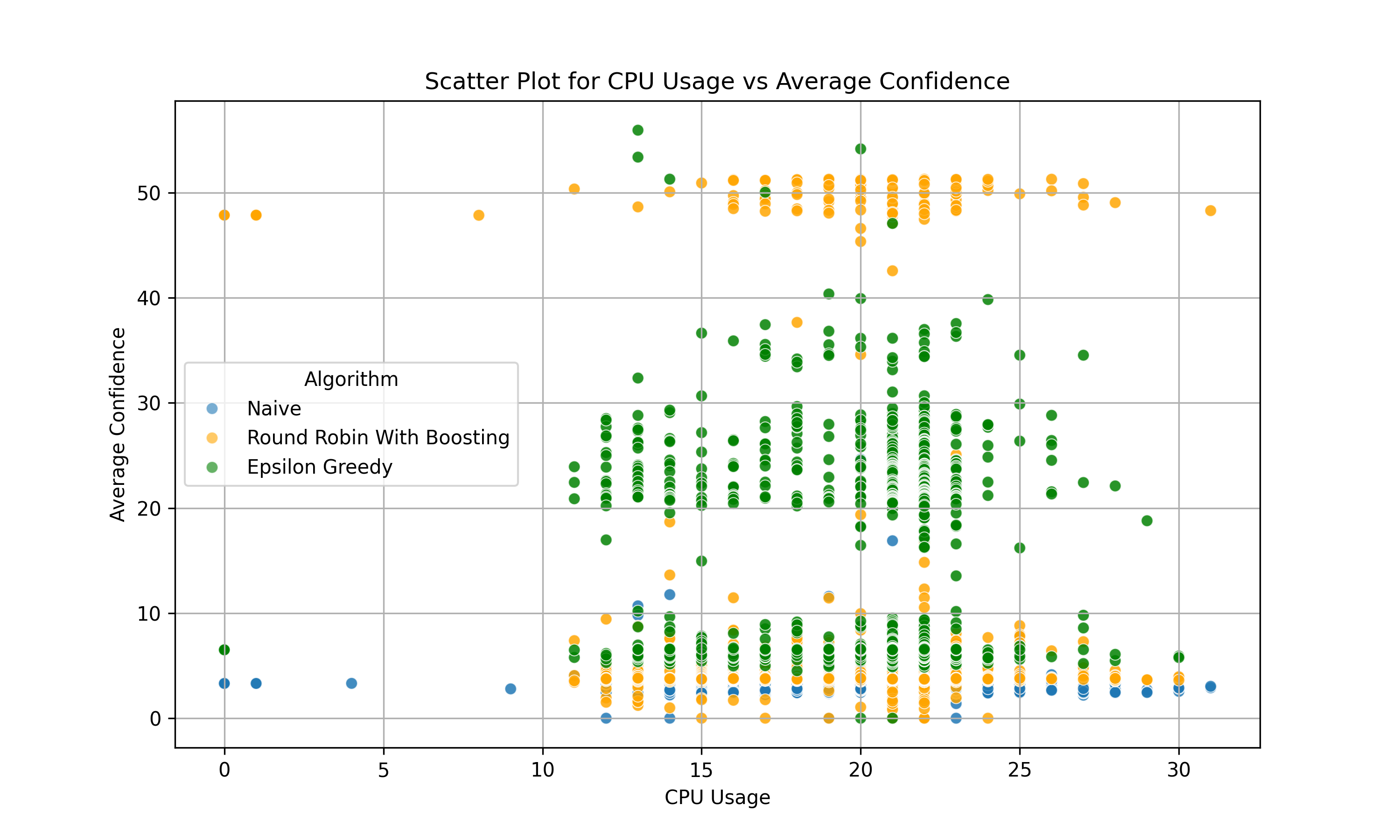}
    \caption{Scatter-plot depicting CPU Usage vs Confidence Score}
    \label{fig:scatter-plot}
\end{figure}

\begin{figure*}[ht]
    \centering
    \includegraphics[width=1\linewidth]{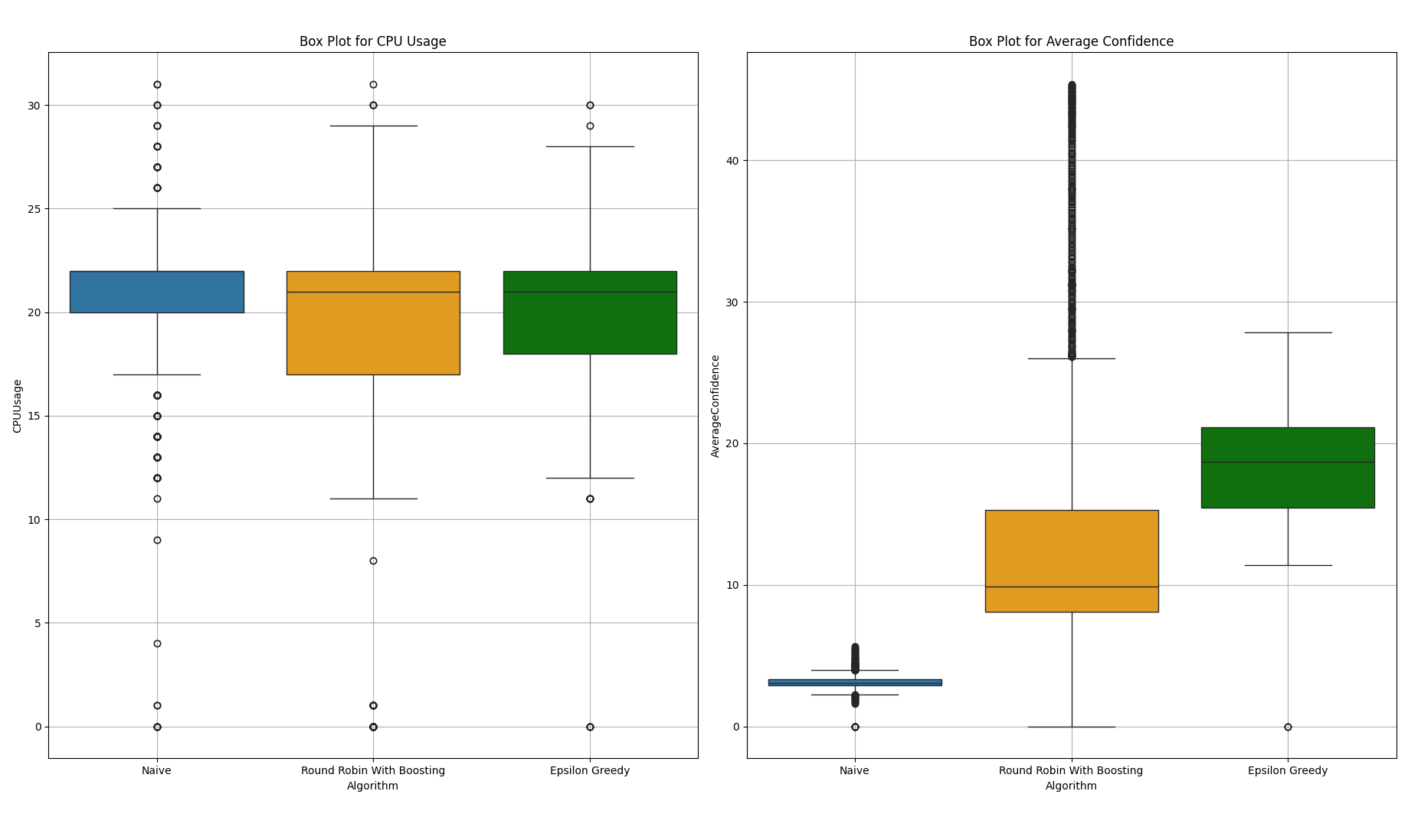}
    \caption{Comparison of All Parameters Using Box-plot}
    \label{fig:box-plot}
\end{figure*}

\begin{figure}
    \centering
    \includegraphics[width=1\linewidth]{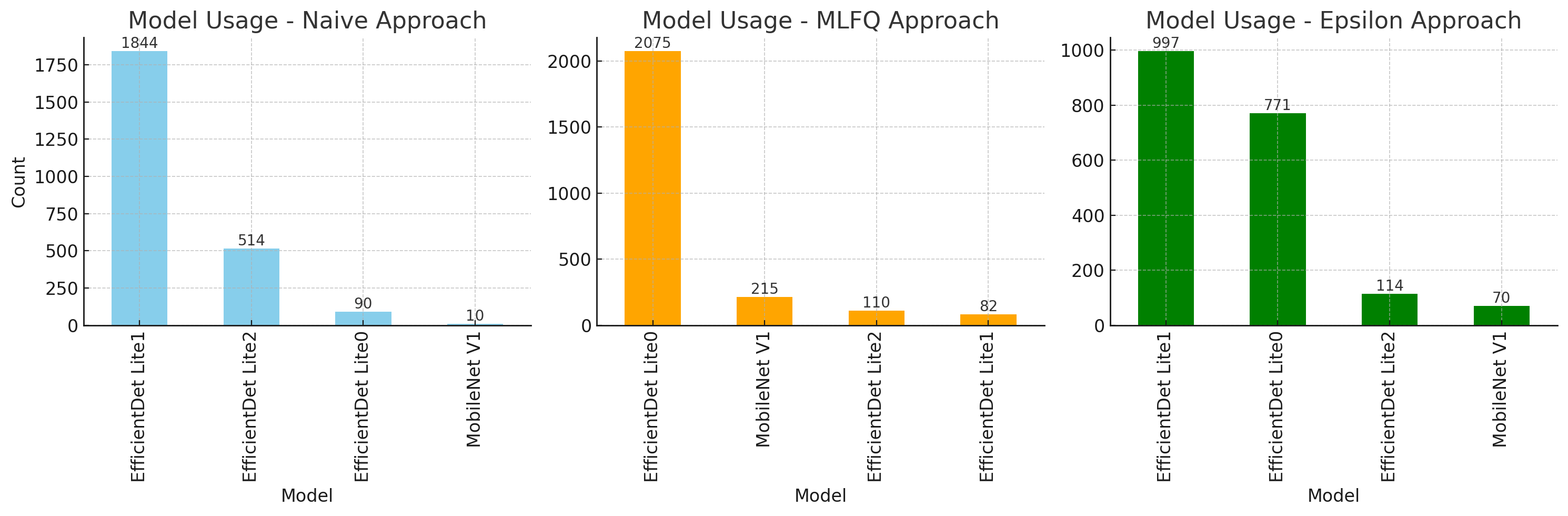}
    \caption{Model Usage for Different Approaches}
    \label{fig:model-usage}
\end{figure}

\begin{figure}
    \centering
    \includegraphics[width=1\linewidth]{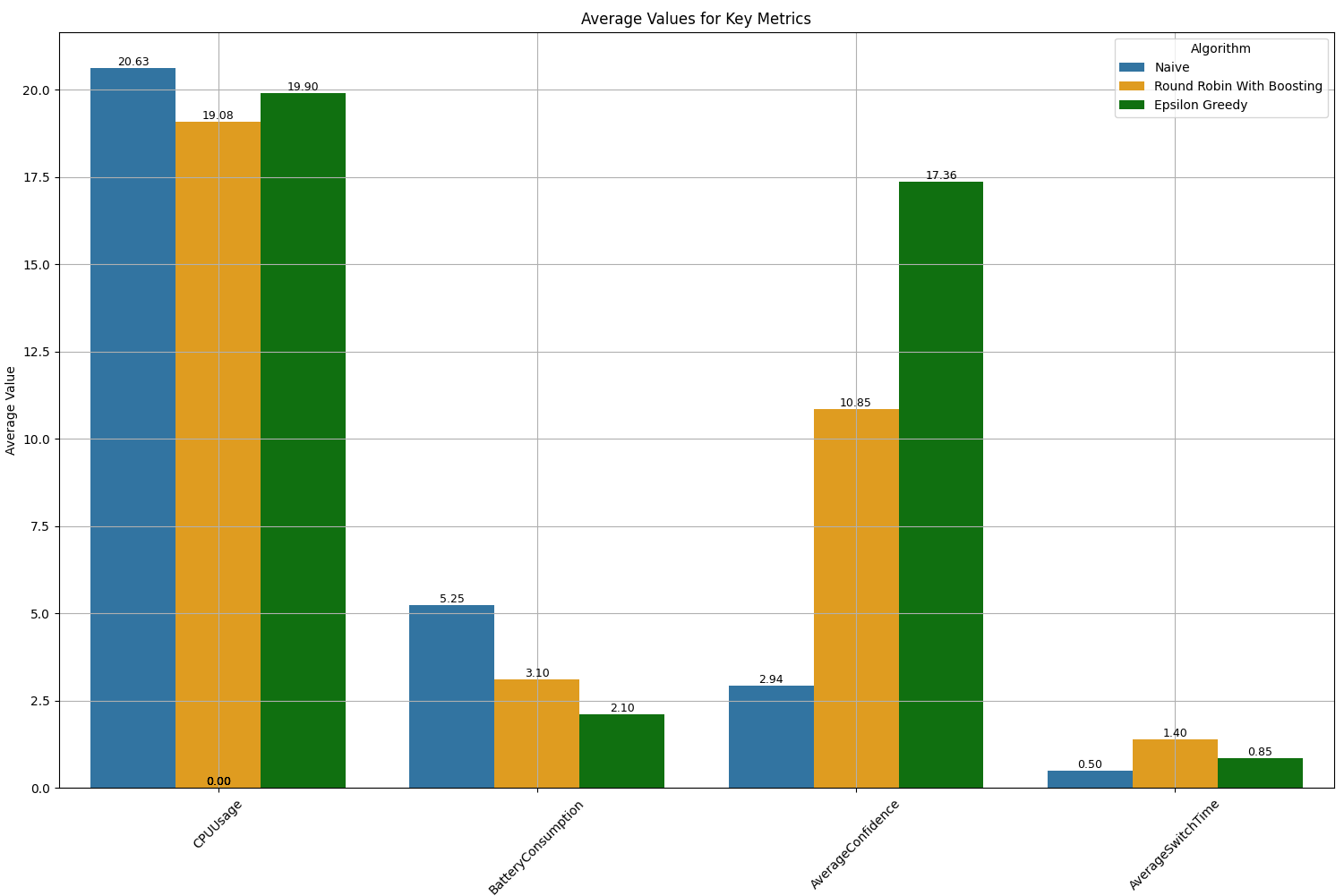}
    \caption{Comparison of Average Values for All Parameters (Including Switch Time)}
    \label{fig:all-parameters-comparision}
\end{figure}


The objective of our EdgeMLBalancer evaluation is to assess the effectiveness, fairness, and efficiency of the approach by answering:

\begin{enumerate}[label=\textbf{RQ\arabic*.}]
    \item How effective is our EdgeMLBalancer approach compared to other approaches in balancing trade-offs between computational efficiency, and detection accuracy within ML-Enabled System?
    \item Which model selection approach ensures the fair allocation of resources between models, effectively balancing detection accuracy and resource utilization, while avoiding model starvation and maintaining robust system performance?
    \item How does the time taken for model switching in our EdgeMLBalancer approach compare to other approaches in terms of its impact on responsiveness and overall system efficiency during real-time operations?
\end{enumerate}

In the remainder of this section, we first discuss our experimental setup, as well as the data used for the evaluation of the approach, following with a discussion of the results for the research questions. 

\subsection{Experimental Setup}

We implemented EdgeMLBalancer as a mobile application, designed to run on Android Devices. The application was developed using Android Studio \cite{b24}, with the primary algorithms written in Kotlin\footnote{\url{https://kotlinlang.org/}}. The TensorFlow Lite library\footnote{\url{https://www.tensorflow.org/resources/libraries-extensions}} was employed for on-device ML model inference, ensuring compatibility with edge-devices. The application integrates \textit{CameraX API}\footnote{\url{https://developer.android.com/media/camera/camerax}} for real-time video feed processing and \textit{MetricLogger} for monitoring and logging system metrics, including CPU usage, and model accuracy. The logging functionality write metrics to CSV file for later analysis. The source code, datasets, and ML models are available here.\footnote{\url{https://github.com/sa4s-serc/EdgeMLBalancer}}

The experiments of EdgeMLBalancer were prototyped using the \textit{Qualcomm QIDK (Qualcomm Innovation Development Kit)}, a platform designed for testing and developing AI applications on edge devices. Equipped with Snapdragon ® 8 Gen 2 processor, Adreno GPU, the QIDK's advanced AI engine and robust connectivity features provided flexibility to simulated various workloads and configurations. The prototyping phase enabled the simulation of various edge scenarios, which helped us to formulate optimization and deployment strategies for \textit{Samsung Galaxy M21} smartphone, on which we tested the primary deployment. 

The \textit{Samsung Galaxy M21} was equipped with an \textit{Exynos 9611 chipset}, featuring an octa-core CPU (4x Cortex-A73 cores clocked at 2.3 GHz and 4x Cortex-A53 cores clocked at 1.7 GHz), 6 GB RAM, and running One UI Core 4.1 based on Android 12. Its 48 MP rear camera (f/2.0) was used to capture experimental video, and its 6000 mAh battery provided sufficient endurance to conduct prolonged experiments without interruptions due to resource-constraints.

To simulate the real scenario of the experiment with as much fidelity as possible, we processed the video in real-time by the application from a 30-minute recording of the Indian-traffic data with 60 frames per second, running continuously throughout the experiment for each approach. This setup was chosen to ensure that the data remained consistent and fair in all approaches, providing a uniform benchmark for evaluation. For the evaluation, we measured different metrics metrics:

\smallskip
\noindent 1. \textit{Accuracy} was calculated as the percentage of correctly detected objects compared to the ground truth. 

\smallskip
\noindent 2. \textit{CPU utilization} of the system (in percentage) during inference while using different models 


\smallskip
\noindent 3. {\em Switching Overhead}, the latency incurred (in ms) while switching between different models when using each of the approaches. 

The approach was evaluated by comparing it against two other baselines resulting in a total of three different experiment candidates. Each of them was executed for a period of 30 minutes, with each run separated by a 30-minute cooldown gap to stabilize the system, preventing any carryover effects. The three experimental candidates are:

\smallskip
\noindent 1. \textit{Naive} approach, switching between models will occur based on the predefined thresholds to balance \textit{accuracy}, and \textit{CPU usage}

\smallskip
\noindent 2. \textit{Round Robin with Boosting} dynamically prioritizes the switching based on time slices and CPU usage, with boosting involving periodic recalibration of CPU usage for all models.

\smallskip
\noindent 3. The proposed \textit{EdgeMLBalancer approach} employed a probabilistic adaptive strategy, using real-time CPU monitoring and workload demands to dynamically select models, optimizing resource usage while maintaining responsiveness.

\subsection{Results}

\noindent
\textbf{RQ1.} \textit{How effective is our EdgeMLBalancer approach compared to other approaches in balancing trade-offs between computational efficiency and detection accuracy within ML-Enabled System?} 

\smallskip
\noindent
We examine our EdgeMLBalancer approach's effectiveness by comparing its performance with two other experiment candidates mentioned in the \textit{experimental setup} of this section, focusing on the balance between CPU usage and inference accuracy. \textit{Table}\ref{tab:table} showcases that Epsilon-Greedy (EdgeMLBalancer) demonstrates effective balance, achieving the highest average accuracy \(17.36\%\) while optimizing resource utilization with average CPU usage \(19.90\%\), despite processing fewer frames (1952). In contrast, the Naive approach processes the highest number of frames (2482), yet this is achieved at the cost of significantly lower average accuracy \(2.94\%\) with the highest average CPU usage \(20.63\%\) , indicating inefficient resource utilization and poor adaptability. However, Round Robin with Boosting processes slightly higher frames (2458) with average accuracy \(10.85\%\) and average CPU usage \(19.08\%\). The \textit{Figure}\ref{fig:scatter-plot} further reinforces these findings. Epsilon-Greedy (EdgeMLBalancer) strikes a balance between CPU usage and accuracy, with an average CPU usage of \(19.90\%\) and a detection accuracy of \(17.36\%\). Compared to Naive and Round Robin approaches, Epsilon-Greedy (EdgeMLBalancer) achieves a \(491.45\%\) (from 2.94\% to 17.36\%) and \(59.94\%\) (from 10.85\% to 17.36\%) higher average accuracy, while reducing average CPU usage by 3.51\% (from 20.63\% to 19.90\%) compared to Naive, and with only a 4.32\% (from 19.08\% to 19.90\%) increase in average CPU usage compared to Round Robin with Boosting. The \textit{Figure}\ref{fig:box-plot} also strengths support our findings that, compared to all approaches, our EdgeMLBalancer approach achieves better performance by effectively managing the trade-offs between computational efficiency and accuracy. 

\smallskip
\noindent
\textbf{RQ2.} \textit{Which model selection approach ensures the fair allocation of resources between models, effectively balancing detection accuracy and resource utilization, while avoiding model starvation and maintaining robust system performance?} 

\smallskip
\noindent
We assess the fairness of approaches by analyzing the distribution of model selection, given that the allocation of resources remains consistent across all approaches, as outlined in the \textit{experimental setup}. \textit{Figure}\ref{fig:model-usage} showcases, that Naive approach selected \textit{EfficientDet Lite1} model 1844 times (74.29\%) out of 2482 frames that are processed, as mentioned in the \textit{Table}\ref{tab:table}. In contrast, Round Robin with Boosting selected \textit{EfficientDet Lite0} model 2075 times (84.41\%) out of 2458 frames. However, Epsilon-Greedy (EdgeMLBalancer) approach selected \textit{EfficientDet Lite1} model 997 times (51.07\%), \textit{EfficientDet Lite0} is selected 771 times (39.49\%), \textit{EfficientDet Lite2} is selected 114 times (5.84\%), and \textit{MobileNet V1} model is selected 70 times (3.58\%) out of 1952 frames. From the \textit{Figure}\ref{fig:model-usage}, we can say that the Epsilon-Greedy (EdgeMLBalancer) represents an improvement in fairness of 43.62\% over Naive and 41.47\% over Round Robin with Boosting in terms of reducing the selection disparity between the models. This is due to the decision-making complexity in approaches like Naive and Round Robin with Boosting, where simplistic or rigid switching mechanisms prioritize certain models without adequately considering runtime conditions or the need for equitable model utilization. Compared to all approaches, our EdgeMLBalancer approach achieves fair distribution among models, reflecting its dynamic adaptability, effectively preventing model starvation. It is important to note that along with guaranteeing fairness, EdgeMLBalancer is also able to balance effectively between CPU Usage and accuracy as demonstrated in the results of RQ1.

\smallskip
\noindent
\textbf{RQ3.} \textit{How does the time taken for model switching in our EdgeMLBalancer approach compare to other approaches in terms of its impact on responsiveness and overall efficiency of the system during real-time operations?}

\smallskip
\noindent
To answer this question, we compare the time taken for model switching in real-time operations of all three approaches. The average switching time of Epsilon-Greedy (EdgeMLBalancer) is 0.85 seconds, as shown in \textit{Figure}\ref{fig:all-parameters-comparision}. This moderate switching time reflects its adaptive model selection mechanism, which evaluates runtime conditions and adjusts models accordingly. In contrast, the Naive and Round Robin with Boosting exhibits the average switching time of 0.50 seconds and 1.40 seconds respectively. While Naive's rapid switching time suggests minimal decision-making complexity, its model usage analysis from results of RQ2 highlights significant model starvation. Similarly, Round Robin with Boosting, takes the longest switching time, paired with model starvation. 

Our approach balanced switching-time, combined with fair model usage, as discussed in \textbf{RQ2}, showcases its superior decision-making mechanism. This underscores the capability of Epsilon-Greedy (EdgeMLBalancer) as the robust and well-rounded approach for real-time operations.

\section{Threats to Validity}

Threats to \textit{external validity} concerns the generalizability of our findings. While our study is prototyped on Qualcomm QIDK platform and tested on a single mobile device as discussed in the experimental setup section, and a specific set of models, the techniques used in our EdgeMLBalancer approach can be extended to other ML tasks and resource-constrained environments with similar challenges, such as energy efficiency, performance, and adaptability. It can also can be extended to other android devices, but non-android devices is beyond the scope of this work.

Threats to \textit{construct validity} concern whether the metrics used accurately represent the phenomena being studied. In our case, potential threats arise from the accuracy of switching time measurements and the model usage fairness calculations. While switching times are measured based on timestamps, minor discrepancies due to system-level latencies might exist. On the other end, the fairness analysis is based on the model usage distribution, which assumes equal utility for all models. This assumption may not hold in scenarios where certain models are inherently more suitable for specific workloads. To address this, the same experimental setup and data sources were used across all approaches to ensure consistent measurement conditions. Additionally, model selection decisions are based on real-time metrics like CPU usage and accuracy, ensuring that the evaluation reflects practical system behavior. 

One potential threat to \textit{internal validity} can be the impact of varying hardware conditions, such as residual background tasks, temperature fluctuations, or battery states, which might affect CPU performance. To mitigate these factors, we took the following precautions: The mobile device was completely reset before testing each approach to eliminate residual processes that could interfere with the experiment. The device was fully charged before starting each approach to ensure uniform initial battery conditions. A warm-up phase was performed before each test to stabilize the hardware state, ensuring consistency throughout the experiments. 

\textit{Conclusion validity} concerns the robustness and reliability of the results. A potential threat is the statistical power of our findings, given the limited duration of experiments (30 minutes per approach). While this duration allowed for meaningful comparisons, it may not fully capture the long-term trends such as battery degradation or model stability under sustained usage, as our decision was to focus only on evaluating the short-term performance and adaptability. To mitigate this, multiple runs were performed to ensure repeatability and comparisons were made using consistent metrics such as CPU usage, accuracy, and switching time. 

\section{Related Work}

The principles of self-adaptation in MLS have been explored in various contexts. Recent advancements, such as Tundo et al. \cite{b28} observed, these systems often rely on predefined configurations, limiting their potential to dynamically adapt to unforeseen runtime conditions. Similarly, while Convolutional Neural Networks (CNNs) have revolutionized object detection \cite{b29}, their application often emphasizes optimizing individual models \cite{b30, b31, b32, b33, b34} rather than system-wide adaptability. 

Notably, a recent survey \cite{b22} highlighted the underutilization of unsupervised learning in self-adaptive systems, echoing the need for scalable, robust architectures for MLS. Despite these advancements, practical applications of self-adaptive systems to real-time scenarios, such as resource-constrained edge-based object detection, remain limited, leaving critical gap in the field. 

The emergence of Green AI has significantly reshaped the landscape of Machine Learning (ML) research, emphasizing energy efficiency and sustainability. Verdecchia et al. \cite{b25} provided a comprehensive review of 98 studies, underscoring a predominant focus on energy efficiency mechanisms within ML systems. However, the translation of these theoretical advancements into practical solutions, particularly in runtime contexts, remains sparse. While early works, such as IBM's autonomic computing vision \cite{b26}, laid the groundwork for self-adaptive systems, these concepts have evolved to encompass ML-enabled systems, introducing the possibility of dynamic adaptability to changing operational conditions.

Multiple studies have explored sustainable development in AI through energy-efficient methods. For instance, Jarvenpaa et al. \cite{b27} identified 12 architectural tactics for sustainability in ML-enabled system, bridging the gap between academic research and practical implementations. Similar studies such as Dagoberto et al. \cite{b11} demonstrated a remarkable 28.7\% reduction in CO2e emissions through hyperparameter optimization, aligning with the Green AI framework. These findings align with broader themes of integrating sustainability-focused tactics at the design stage to achieve computational efficiency and reduce ecological footprints. 

Efforts like those by Dagoberto et al. \cite{b12} studied AutoML systems, optimizing energy consumption without comprising accuracy. These studies contribute to the ongoing development of eco-conscious ML processes by introducing metrics and practices that prioritize energy efficiency during training and inference phases. However, while these approaches address sustainability at the design level, runtime energy efficiency, particularly for real-time applications, remains under-explored.

Departing from the aforementioned approaches, our work focuses on real-time object detection at the edge, leveraging runtime adaptability. It is based on a model selection approach grounded in Epsilon-Greedy exploration to optimize real-time object detection at the edge. Further as demonstrated by our results, this approach balances computational efficiency, accuracy, and resource fairness by dynamically adapting to runtime conditions. 

\section{Conclusions and Future Work}

 By balancing model-switching using computational efficiency, accuracy, and resource fairness, the approach addresses key challenges in edge AI systems, including adaptability and sustainability. Our findings highlight significant improvements in model usage fairness, computational efficiency, and detection accuracy compared to other approaches. 
Further, the results validate the potential of adaptive strategies to optimize edge-AI systems for compute-conscious, high-performance applications. We believe that our study shows promising results and forms the basis for future exploration to switch between models on edge devices. However, further studies are needed, which will be part of our future work.

Building upon this study, future work includes but is not limited to: (i) integrating Large Language Models (LLMs) to complement vision-based tasks (already in the pipeline), (ii) extending it to other Android devices and other edge devices like raspberry pi 4, Nvidia Jetson nano etc., and (iii) then extending the approach to hybrid edge-cloud architectures would enable optimized workload distribution between edge devices and cloud infrastructure, balancing real-time constraints with computational efficiency.

\section{Acknowledgements}


We sincerely thank Qualcomm Inc. for their generous funding and support for this study through the Qualcomm EdgeAI labs at the IIIT Hyderabad, India. Their contribution has been instrumental in enabling the research and development of this work. We further thank Arya Marda from Software Engineering Research Center, IIIT Hyderabad, India, in supporting us at the time of development.

\end{document}